Kasirga Yildirak[1], Osman Gulseven[2]

# INDEMNITY PAYMENTS IN AGRICULTURAL INSURANCE: RISK EXPOSURE OF EU STATES*

*This study estimates the risk contributions of individual European countries regarding the indemnity payments in agricultural insurance. We model the total risk exposure as an insurance portfolio where each country is unique in terms of its risk characteristics. The data has been collected from the recent surveys conducted by the European Commission and the World Bank. Farm Accountancy Data Network is used as well. 22 out of 26 member states are included in the study. The results suggest that the Euro-Mediterranean countries are the major risk contributors. These countries not only have the highest expected loss but also high volatility of indemnity payments. Nordic countries have the lowest indemnity payments and risk exposure.*

Касірга Їлдірак, Осман Гюльсевен

# КОМПЕНСАЦІЙНІ ВИПЛАТИ У СІЛЬСЬКОГОСПОДАРСЬКОМУ СТРАХУВАННІ: СХИЛЬНІСТЬ ДО РИЗИКІВ РІЗНИХ КРАЇН-ЧЛЕНІВ ЄС

*У статті оцінено ризики окремих європейських крän щодо компенсаційних виплат у сільськогосподарському страхуванні. Змодельовано сумарний ризик страхового портфелю для кожної країни, ризикові особливості яких є унікальними. Дані для моделювання отримано з останніх звітів Єврокомісії, Світового банку та Мережі фермерського бухобліку. Використано дані по 22 країнах ЄС з 26. Результати дослідження показали, що до найбільших ризиків схильні середземноморські країни ЄС. Цим країнам притаманні високі очікувані збитки та нестабільність компенсаційних виплат. Північні країни ЄС мають найнижчі показники — як по компенсаційних виплатах, так і по схильності до ризику.*

***Ключові слова:*** *Єдина сільськогосподарська політика ЄС; сільськогосподарське страхування; схильність до ризиків; компенсаційні виплати; страховий портфель.*
*Форм. 8. Табл. 3. Літ. 12.*

Касирга Илдирак, Осман Гюльсевен

# КОМПЕНСАЦИОННЫЕ ВЫПЛАТЫ В СЕЛЬСКОХОЗЯЙСТВЕННОМ СТРАХОВАНИИ: ПОДВЕРЖЕННОСТЬ РИСКАМ РАЗНЫХ СТРАН-ЧЛЕНОВ ЕС

*В статье оценены риски отдельных европейских стран касательно компенсационных выплат в сельскохозяйственном страховании. Смоделирован суммарный риск страхового портфеля для каждой страны, рисковые особенности которых уникальны. Данные для моделирования получены из последних отчетов Еврокомиссии, Мирового банка и Сети фермерского бухучета. Использованы данные по 22 странам ЕС из 26. Результаты исследования показали, что наибольшим рискам подтверждены средиземноморские страны*

---

[1] Assistant Professor of Economics, Trakya University, Edirne, Turkey.
[2] Assistant Professor of Economics and Marie Curie Fellow, Middle East Technical University, Ankara, Turkey.
* The AgInsurance Project is supported by European Commission under Marie Curie Grant #247723. The opinions presented in this research are those of the researchers, and do not necessarily reflect the opinion of the Commission. http://www.aginsuranceproject.com

*ЕС. Этим странам характерны высокие ожидаемые убытки и нестабильность компенсационным выплат. Северные страны ЕС имеют наименьшие показатели как по компенсационным выплатам, так и по подверженности риску.*

*Ключевые слова:* Единая сельскохозяйственная политика ЕС; сельскохозяйственное страхование; подверженность рискам; компенсационные выплаты; страховой портфель.

**Introduction.** Farmers face unavoidable risks in agriculture, due to a variety of reasons. The risk factors in agriculture are inevitable, but they are manageable to a large extent, through risk management strategies. Agricultural risks can be classified into 5 categories: climatic, health-related, geological, market-related and human (Wenner and Arias, 2003).

In the late 1980s and early 1990s, the insurance industry started to become interested in natural disasters since weather events create opportunities for portfolio diversification due to the fact that they are unrelated with financial markets, big catastrophes may be easily financed through financial markets; government can also play a regulatory role, and investors would be able to choose a particular type of risk unlike direct investment in an insurance company (Miranda and Vedenov, 2001). Agricultural insurance has many advantages in agricultural risk management. However, it has the disadvantage of suffering from market incompletion in the form of information asymmetries or incomplete contingent claims markets. Dismukes et al. (2004) develop a measure in valuing agricultural insurance, by assuming only that rational producers never forego opportunities to risklessly raise profit. Wenner and Arias (2003) find that the necessary conditions for the development of innovative instruments are a suitable technology, an institutional support system, means of institutional delivery and a proper legal and regulatory framework for insurance company supervision.

Government involvement in agriculture is justified by the market failures explained above. In analyzing the problems of market insurance in agriculture, Goodwin (2001) argues that crop insurance is a major kind of income subsidy for farmers and a major step in providing economic support to the agricultural sector. However, the problems of moral hazard and adverse selection make it impossible in the agricultural insurance sector to provide private insurance contracts, as well as the existence of systemic risk due to weather events which create correlation in the risks of individual farmers. Subsidy for agricultural insurance is necessary under these circumstances (Kailiang and Wenjun, 2007).

There are many types of agricultural insurance. The first one is the single-risk insurance, which covers against one risk only. Combined (peril) insurance covers multiple risks. Price insurance covers against price decreases under some threshold, whereas yield insurance covers production risks. Revenue insurance combines the price and the yield insurances. Other types of agricultural insurance are whole-farm insurance, income insurance, and index insurance (European Commission, 2008). More insurance products and improving the risk assessment methods have the potential to increase the incentives of farmers to participate in the agricultural insurance programs. Makki and Somwaru (2001) propose 4 new tools, namely adjusted gross revenue insurance which is whole-farm rather than on a crop-by-crop basis; tax-deferred savings accounts for farmers; area revenue insurance – Group Risk Income

Protection (GRIP) which is more attractive than GRP due to its income component; and regional weather index insurance which can effectively reduce asymmetric information and facilitate the development of an independent crop insurance market. Teaching farmers about the benefits of agricultural insurance is also important in increasing their participation in the market. For example, Carter et al. (2009) use a financial education game to teach how to use index insurance in 5 different communities in Kenya, and find that the farmers' participation has increased after the game series.

Concerning the valuation of agricultural insurance, Chambers (2007) provides a model for its valuation. The basic assumption of his model is that rational producers never forego opportunities to risklessly raise profit. The model estimates a stochastic discount factor for the farmers that are independent of their risk preferences. Mahul (2001a, 2001b) proposes an insurance contract that is optimal for the climatic risks under the uninsurable and dependent aggregate production risk.

In our study we introduce a totally different approach that will close the gap left by the revious studies. We consider agricultural insurance as a financial derivative option to manage risks faced by European farmers. We estimate the risk contributions of individual European countries regarding the indemnity payments in agricultural insurance; using data from the recent surveys conducted by the European Commission and the World Bank, as well as the Farm Accountancy Data Network (FADN).

**Data and Methodology.** The data is collected from various resources. There is no standardized data on the indemnity payments to agricultural insurance. Partial information on insurance data is derived from the EU funded research reports. (European Commission, 2008). This data is also compared with 2008 Survey conducted by the World Bank. Data on 4 countries were either inconsistent or nonexistent. Therefore we removed these countries (Belgium, Ireland, Latvia, and Slovakia). Microlevel Farm Accountancy Data Network (FADN) data was also used to estimate the variability of production. There are 22 countries included in our study. These member states share 96% of the agricultural production revenues in the European Union.

Fluctuations in crop and livestock production that are not included in the existing insurance systems are beyond our research. Our definition of risk exposure is based on the current insured value for each country. Based on this definition the mean risk exposure for each country is the same as average indemnity payments to the insured farmers. The standard deviation measure the variability of the indemnity payments. However, the data on indemnity payment variability for each country does not exist. Therefore, we defined a proxy parameter based on the deviation of production from the trend. Table I lists the mean and standard deviation of risk exposures for each country included in our estimation.

As expected, large industrialized countries with significant agricultural production have much higher risk exposure than other member states. Germany and France, dual-dynamos of the Union, have risk exposures worth 13.5 bln. Euros and 11.7 bln. Euros respectively. On the contrary, the greatest expected losses arise from the Mediterranean states, namely, Greece, Spain and Italy. Spain has the highest indemnity payments of 370 million Euros, followed by Greece (218 mln. Euros), and Italy (208 mln. Euros).

*Table I.* **Country Level Risk Parameters**

| Country | # | Risk Exposure (Million €) | Mean Risk | Std. Dev. | CROP Ratio | Livestock Ratio | Expected Loss (Million €) |
|---|---|---|---|---|---|---|---|
| (BGR) Bulgaria | 1 | 800,12 | 3,12% | 0,72% | 0,65 | 0,35 | 24,96 |
| (CYP) Cyprus | 2 | 328,82 | 6,84% | 3,40% | 0,49 | 0,51 | 22,49 |
| (CZE) Czech Rep. | 3 | 1665,14 | 1,31% | 0,35% | 0,56 | 0,44 | 21,87 |
| (DAN) Denmark | 4 | 6577,06 | 0,65% | 3,84% | 0,34 | 0,66 | 42,75 |
| (DEU) Germany | 5 | 13585,30 | 1,00% | 1,34% | 0,44 | 0,56 | 135,30 |
| (ELL) Greece | 6 | 9436,51 | 2,31% | 1,22% | 0,71 | 0,26 | 218,00 |
| (ESP) Spain | 7 | 8494,09 | 4,35% | 2,48% | 0,64 | 0,36 | 369,23 |
| (EST) Estonia | 8 | 160,20 | 0,65% | 1,17% | 0,44 | 0,56 | 1,04 |
| (FRA) France | 9 | 11762,32 | 1,11% | 0,48% | 0,54 | 0,46 | 129,97 |
| (HUN) Hungary | 10 | 3382,78 | 0,96% | 3,54% | 0,60 | 0,10 | 32,62 |
| (ITA) Italy | 11 | 4332,43 | 4,81% | 3,32% | 0,61 | 0,39 | 208,39 |
| (LTU) Lithuania | 12 | 86,73 | 4,30% | 2,67% | 0,26 | 0,74 | 3,72 |
| (LUX) Luxembourg | 13 | 86,48 | 1,98% | 1,83% | 0,47 | 0,53 | 1,71 |
| (NED) Netherlands | 14 | 7327,41 | 1,44% | 2,41% | 0,47 | 0,53 | 105,75 |
| (OST) Austria | 15 | 3238,08 | 1,87% | 1,50% | 0,29 | 0,71 | 60,61 |
| (POL) Poland | 16 | 1507,09 | 1,33% | 1,88% | 0,51 | 0,49 | 20,09 |
| (POR) Portugal | 17 | 583,87 | 5,04% | 2,93% | 0,55 | 0,45 | 29,42 |
| (ROU) Romania | 18 | 1108,24 | 1,26% | 0,44% | 0,45 | 0,55 | 14,00 |
| (SUO) Finland | 19 | 35,39 | 2,26% | 0,83% | 0,37 | 0,63 | 0,80 |
| (SVE) Sweden | 20 | 2524,72 | 2,52% | 0,35% | 0,41 | 0,59 | 63,74 |
| (SVN) Slovenia | 21 | 146,99 | 7,25% | 2,84% | 0,47 | 0,53 | 10,65 |
| (UKI) UK | 22 | 1398,33 | 0,56% | 1,16% | 0,44 | 0,60 | 7,83 |
| Total (22 EU States) | | 78568,21 | 2,28% | 1,85% | 0,51 | 0,49 | 1525,03 |

Model:

We consider European agricultural insurance exposure as a portfolio of credit loan with default probabilities that differ for each member state. Our model is a Poisson-Gamma model also known as CreditRisk+ developed by CreditSuisse™. Let $A_i$, $i = 1,2,...I$ be the $i^{th}$ obligor with rating $R_s$, $s = 1, 2,..., S$ and $X_{i,t}$ be the total amount (coverage) on the contract that will be paid to obligor at time $t = 1, 2,...,T$, with $T$ representing maturity of a contract. Let $x_i = e^{-r\tau} X_{i,t}$ the present value of a coverage. $r$ represents the interest rate and $\tau$ is the annual time to termination. Let's assume $x_i$ has been lent to the $i^{th}$ insured country, and it will be paying it back by regular payments. Let us introduce the probability generating function for $i$ defined in terms of an auxiliary variable $z$ by

$$F_i(z) = \sum_{n=0}^{\infty} P(n \, defaults) z^n = 1 - P_i + P_i z \qquad (1)$$

where $P_i$ is the default probability of $i^{th}$ state. We can rewrite the above equation as

$$F(z) = \prod_{i=1}^{I} F_i(z) = e^{\mu(z-1)} = \sum_n \frac{e^{-\mu} \mu^n}{n!} z^n \qquad (2)$$

where $\mu = \sum_i P_i$.

Using Poisson distribution we model the probability of having $n$ defaults in a portfolio. Severity of defaults is also introduced to the model. This is done by map-ping exposures into corresponding integer valued exposure bands so that probability generating function technique could be used.

Then exposure band $v_i$ is computed as $v_i = $ *roundup [max $x_i/L$]* where ceil represents rounding toward positive infinity and $L$ is some suitable size of unit that normalizes exposures. Then $v_i$ is the nearest upper integer to the normalized exposure. Let $\lambda_i$ be the expected loss and $\varepsilon_i$ be the expected loss expressed as multiples of the unit $L$. For a sufficiently large portfolio, more than one obligor shares a common band. Let $v_j$ be the j$^{th}$ exposure band, $\varepsilon_j$ be the expected loss in units of $L$ and $\mu_j$ be the number of defaults in band $j$. It follows that

$$\mu_j = \frac{\varepsilon_j}{v_j} = \sum_{i:v_i=v_j} \frac{\varepsilon_i}{v_i}. \qquad (3)$$

Then total expected number of defaults in the entire portfolio with $m$ exposure bands will be:

$$\mu = \sum_{j=1}^{m} \frac{\varepsilon_j}{v_j}$$

Using this definition we can form the probability generating function for the case where severity of defaults is taken into consideration:

$$G(z) = p(n*Loss)z^n = \prod_{j=1}^{m} G_j(z) \qquad (4)$$

where $n*Loss$ is the aggregate loss. Explicit form for $G(z)$ is written as

$$G(z) = e^{-\sum_{j=1}^{m}\mu_j + \sum_{j=1}^{m}\mu_j z^{v_j}} \qquad (5)$$

In order to numerically estimate the aggregate loss function, one can introduce the polynomial

$$P(z) = \frac{\sum_{j=1}^{m} \left(\frac{\varepsilon_j}{v_j}\right) z^{v_j}}{\sum_{j=1}^{m} \left(\frac{\varepsilon_j}{v_j}\right)} \qquad (6)$$

We can write probability generating function in a more compact form as $G(z) = F(P(z))$, where $F$ is derived from equation (2).

In the insurance portfolio there are $K$ different sectors. Let us introduce $y_k$ representing the average default rate over the sector $k$. Assume that the default rates in sectors follow gamma distribution with parameters $\alpha_k$ and $\beta_k$ with:

$$\alpha_k = \frac{\mu_k^2}{\sigma_k^2} \quad \text{and} \quad \beta_k = \frac{\sigma_k^2}{\mu_k},$$

where $\mu_k$ is the mean of $y_k$ and $\sigma_k$ is the standard deviation of $y_k$. The probability generating function for default events from the whole portfolio is given by

$$F(z) = \prod_{k=1}^{K} F_k(z) = \prod_{k=1}^{K} \left(\frac{1-p_k}{1-p_k z}\right)^{\alpha_k} \qquad (7)$$

where $p_k = \dfrac{\beta_k}{1+\beta_k}$

Then the distribution of default losses is obtained as

$$G(z) = \prod_{k=1}^{K} G_k(z) = \prod_{k=1}^{K} \left[ \frac{1-p_k}{1-\dfrac{p_k}{\mu_k}\sum_{j=1}^{m(k)}\dfrac{\varepsilon_j^k}{v_j^k}z^{v_j^k}} \right]^{\alpha_k} \quad (8)$$

We compute this function in MATLAB using a combination of Fast Fourier Transformation and Panjer recursions (Nazliben, K. and Yildirak, K.).

**Results:**

In order to estimate the insurance payments, we used the latest MATLAB module to run Creditrisk+ algorithm to estimate the credit default risk. The inputs were risk factors and insured value. The simulations returned the estimated indemnity payments given specific probabilities. Table II shows the simulation outcomes: we observe a strong positive and exponential correlation between probability distribution and estimated indemnity payments. While the mean value indemnity payments equal 1.525 bln. Euros, there is a 1% probability that indemnity payments might be as high as 14.389 bln. Euros.

*Table II.* **Marginal Percentiles**

| Probability | Indemnity Payments (mln. €) |
|---|---|
| 0.5000 | 1525,30 |
| 0.1000 | 6913,91 |
| 0.0500 | 9424,59 |
| 0.0250 | 11824,41 |
| 0.0100 | 14389,19 |
| 0.0050 | 17888,45 |
| 0.0025 | 20418,58 |
| 0.0010 | 23259,33 |

Table III dissects the individual risk contributions of each member state for specific probability levels:

*Table III.* **Country Level Probability Specific Risk Contributions**

| Country | Std. Dev. | Expected Loss (mln. €) | P(A) = 0.1 (mln. €) | P(A) = 0.05 (mln. €) | P(A) = 0.01 (mln. €) |
|---|---|---|---|---|---|
| (BGR) Bulgaria | 0,72% | 24,96 | 39,07 | 45,64 | 58,64 |
| (CYP) Cyprus | 3,40% | 22,44 | 44,71 | 37,84 | 47,52 |
| (CZE) Czech Rep. | 0,35% | 21,81 | 46,71 | 55,38 | 76,48 |
| (DAN) Denmark | 3,84% | 42,75 | 180,66 | 244,91 | 371,97 |
| (DEU) Germany | 1,34% | 135,85 | 992,92 | 1392,26 | 2181,90 |
| (ELL) Greece | 1,22% | 217,97 | 1202,09 | 1660,61 | 2567,30 |
| (ESP) Spain | 2,48% | 369,49 | 1877,31 | 2579,84 | 3969,03 |
| (EST) Estonia | 1,17% | 1,04 | 1,6 | 1,85 | 2,37 |
| (FRA) France | 0,48% | 130,56 | 854,21 | 1191,38 | 1858,09 |
| (HUN) Hungary | 3,54% | 32,47 | 88,70 | 114,91 | 166,72 |
| (ITA) Italy | 3,32% | 208,37 | 658,61 | 868,39 | 1283,21 |
| (LTU) Lithuania | 2,67% | 3,70 | 5,16 | 5,84 | 7,18 |
| (LUX) Luxembourg | 1,83% | 1,70 | 2,39 | 2,70 | 3,33 |

**The End of** *Table 3*

| Country | Std. Dev. | Expected Loss (mln. €) | P(A) = 0.1 (mln. €) | P(A) = 0.05 (mln. €) | P(A) = 0.01 (mln. €) |
|---|---|---|---|---|---|
| (NED) Netherlands | 2,41% | 105,51 | 474,79 | 646,85 | 987,08 |
| (OST) Austria | 1,50% | 60,55 | 161,58 | 208,65 | 301,73 |
| (POL) Poland | 1,88% | 20,04 | 40,16 | 49,54 | 68,07 |
| (POR) Portugal | 2,93% | 29,38 | 46,52 | 54,50 | 70,28 |
| (ROU) Romania | 0,44% | 13,96 | 25,41 | 30,74 | 41,28 |
| (SUO) Finland | 0,83% | 0,79 | 1,56 | 1,92 | 2,64 |
| (SVE) Sweden | 0,35% | 63,60 | 151,67 | 192,70 | 273,83 |
| (SVN) Slovenia | 2,84% | 10,58 | 16,80 | 19,69 | 25,42 |
| (UKI) U.K. | 1,16% | 7,83 | 15,07 | 18,44 | 25,11 |
| TOTAL | 1,85% | 1525,03 | 6913,91 | 9424,59 | 14389,19 |

While expected losses are within reasonable levels, the volatility of indemnity payments is a strong warning signal for EU policymakers. The high level of variability in the Mediterranean countries shows weakness of these countries against large scale shocks. Although, the expected indemnity payment for Greece is around 217 mln. Euros, there is a 1% possibility, that it could be as much as 10 times higher, above 2.5 bln. Euros. A similar risk exists for Spain and Italy. The expected indemnity payments for these countries are around 370 mln. Euros and 208 mln. Euros, respectively. However, there is a 1% possibility that the indemnity payments can be as high as 4 bln. Euros for Spain, and 1.3 bln. Euros for Italy.

**Conclusion and Policy Implications.** In this paper, we applied an innovative methodology to estimate the agribusiness related insurance risk, and indemnity payments. Our results supported the previous studies on the tremendous variation in agricultural risks in the EU member states. While large industrialized countries have the highest level of risk exposure, the southern states have the most volatile risk exposures. The volatility regarding indemnity payments can be attributed mainly to two factors, namely climate variation, and government policy on agricultural insurance.

It is known that Mediterranean region has the most volatile climate. The risk of draught in harvest season is highest in Spain, Portugal, Italy and Greece. The statistical analysis in this paper suggests that farmers in these countries face more risks than in the northern states of the EU. As the effects of global warming gets increasingly evident, agricultural businesses in these states might suffer the most. Consequently, a significant portion of these losses will be reflected in the insurer's balance sheets, as well as in state budgets.

The existing insurance subsidy scheme in the EU's Mediterranean members is another significant factor that contributes to exponentially increasing indemnity payments. These countries provide heavily subsidized insurance options to protect their farmers against weather and disease risks. While, this is a great benefit to farmers, it significantly increases the indemnity payment risk exposure of these countries. The volatility of yields is pretty high in Greek agribusinesses. As we show in our results, there is a 1% possibility of paying more than 2.5 bln. Euros in indemnity payments. Greek government – known for its inefficient institutional settings – almost completely dominates the insurance market through Hellenic Agricultural Insurance (ELGA). The government also enforces the agricultural insurance as a publicly pro-vided compulsory service, leaving no room for private companies. Maybe, it is time to reconsider this policy given Greece's current debt problem and huge budget deficit.

Стаття надійшла до редакції 05.05.2011.